\documentclass[a4paper,12pt]{article}

\usepackage{a4wide}
\usepackage{amssymb}
\usepackage{amsmath}
\usepackage[dvips]{graphicx} 
\usepackage{dcolumn} 
\usepackage{bm} 

\DeclareFontFamily{U}{rsfs}{\skewchar\font"7F}
\DeclareFontShape{U}{rsfs}{m}{n}{%
    <-6> rsfs5
    <6-8> rsfs7
    <8-> rsfs10
    }{}
\DeclareMathAlphabet{\mathscr}{U}{rsfs}{m}{n}

\renewcommand{\ell}{\ensuremath{l}}

\begin{document}

\hfill IFT/4/2006 (Warsaw Univ.)\\

\begin{center}
{\Large\sl
Rotating Black Holes at Future Colliders.\\
III. Determination of Black Hole Evolution \\
}
\vspace{0.5cm}
  {\sc Daisuke Ida${}^1$, Kin-ya Oda${}^2$, and Seong Chan Park${}^3$}\\
  \bigskip
\end{center}
\begin{flushleft}
  {${}^1$Department of Physics, Gakushuin University,
  Tokyo 171-8588, Japan} \\
\vspace{0.25cm}
  {${}^2$Institute of Theoretical Physics, Warsaw University,
  Ho\.za 69, Warsaw 00-681, Poland} \\
\vspace{0.25cm}
  {${}^3$LEPP, Cornell University,
  Ithaca, N.Y.\ 14853, U.S.A.}\\
\end{flushleft}
  \bigskip
\begin{center}
  \textbf{\large Abstract}
  \end{center}
\noindent
TeV scale gravity scenario predicts that the black hole production dominates over all other interactions above the scale and that the Large Hadron Collider will be a black hole factory.
Such higher dimensional black holes mainly decay into the standard model fields via the Hawking radiation whose spectrum can be computed from the greybody factor.
Here we complete the series of our work by showing the greybody factors and the resultant spectra for the brane localized spinor and vector field emissions for arbitrary frequencies.
Combining these results with the previous works, we determine the complete radiation spectra and the subsequent time evolution of the black hole.
We find that, for a typical event, well more than half a black hole mass is emitted when the hole is still highly rotating, confirming our previous claim that 
it is important to take into account the angular momentum of black holes.

\newpage
\section{Introduction}
Black hole has been playing crucial roles for decades in the yet unaccomplished theoretical development to reconcile gravitational interactions with quantum description of
nature~\cite{Bekenstein:1973ur,Hawking:1974sw,Hawking:1976de,Giddings:1995gd,Strominger:1996sh,Horowitz:1996nw,Damour:1999aw}.
From this point of view, it is no doubt desirable to have a direct experimental test of the Hawking radiation~\cite{Hawking:1974sw}
through which black hole is expected to radiate off quanta 
almost thermally.
An 
astrophysical
black hole is too big to have a sizable Hawking temperature for this purpose.
It has long been 
known
that a collision of two particles above the Planck energy scale inevitably leads to a production of
black holes~\cite{'tHooft:1987rb}, which can have 
large Hawking temperature.

Until recently, the Planck scale 
was always assumed to be unaccessibly high for human beings even in theories with extra compact dimensions such as string theory, under the assumption that the compactification radius is of the order of the Planck length.
Without prejudice, there is no reason that forces one to keep this assumption.
Recently it has been pointed out that
the large extra dimensions can reduce the higher dimensional Planck scale down to TeV scale to solve the so-called hierarchy problem without contradicting any experimental data if the standard model interactions are confined to a 3-brane~\cite{Arkani-Hamed:1998rs,Antoniadis:1998ig,Arkani-Hamed:1998nn}.
 It has also been pointed out that a TeV scale gravity is realized if an extra dimension is compactified with the warped metric on $AdS_5$~\cite{Randall:1999ee}.

The TeV scale gravity opens up a possibility of producing 
black holes
and observing 
their decay 
products
directly at next generation collider.
First it was considered that such a black hole will mainly decay into the bulk graviton modes~\cite{Argyres:1998qn,Banks:1999gd}.
Later it has been noticed that the decay channels into observable brane-localized standard model fields dominate over the bulk graviton emission~\cite{Emparan:2000rs}.
When the black hole is highly rotating, the bulk graviton emission is expected to be greatly enhanced~\cite{Frolov:2002as,Frolov:2002gf,Cavaglia:2003hg,Stojkovic:2004hp}.
Currently the bulk graviton equation is obtained only for the non-rotating $J=0$ hole~\cite{Cornell:2005ux,Cardoso:2005mh,Creek:2006ia} and we cannot conclude how much the graviton emission will be enhanced.
There is also an argument that the higher dimensional graviton has a larger number of modes than four dimensions and that the graviton emission would be enhanced due to this fact~\cite{Frolov:2002as,Frolov:2002gf}. The actual number of modes that have smaller masses than the Hawking temperature, which is always smaller than the higher dimensional Planck scale, will depend on the details of the moduli stabilization. Typically one has much fewer light modes after the moduli fixing.
The possibility of black hole production at collider and the detection of its Hawking radiation is studied from general perspective in \cite{Giddings:2001bu} and the experimental signature at the CERN Large Hadron Collider (LHC) is studied in \cite{Dimopoulos:2001hw} in the approximation that the black hole has vanishing anglar momentum $J=0$ and that the Hawking radiation spectrum is described with the high frequency limit $r_h\omega \gg1$, where $r_h$ and $\omega$ are the black hole horizon radius and the energy of the emitted particle, respectively.

In \cite{Ida:2002ez}, we have emphasized the importance to take into account the angular momentum of the black hole and have first estimated the differential production cross section for a given $J$,
whose integral over $J$ qualitatively agrees with the numerical results of the classical gravitational collision of  two massless particles~\cite{Yoshino:2002tx,Yoshino:2005hi}, namely, the fact that the total cross section increases from the Schwarzschild approximation more and more for higher dimensions is well described.\footnote{
A similar formula has appeared in \cite{Anchordoqui:2003jr}, which predicts that the cross section would be reduced by taking into account the angular momentum, contrary to our result. See also \cite{Park:2001xc} for another earlier attempt.}
Further, utilizing the
Newman-Penrose formalism~\cite{Penrose:1985jw}
we have obtained the field equations of the brane-localized scalar ($s=0$), spinor ($s=1/2$), and vector ($s=1$) fields in a separable form for the $D=4+n$ dimensional black hole background with a single angular momentum $J\geq0$~\cite{Myers:1986un}.
Later these equations are also derived for a non-rotating $J=0$ black hole in Refs.~\cite{Kanti:2002ge,Harris:2003eg}.

Once shown to be separable, its radial part can be solved to obtain the greybody factors that completely determine the Hawking radiation spectra without relying on the high frequency limit $r_h\omega\gg1$.
A thermal radiation spectrum at the horizon is modified  when the radiation passes through the curved geometry toward an observer located at the spatial infinity.
The greybody factor is introduced to take this correction into account
for each angular mode $(\ell,m)$ and frequency $\omega$ of the emitted radiation.
It is defined to be the absorption rate by the hole for a steady in-falling flux at the infinity
\begin{align}
  \Gamma_{\ell m}
    &= {
\dot{N}_{\ell m}^{\mathrm{(in)}}-\dot{N}_{\ell m}^{\mathrm{(out)} }
\over \dot{N}_{\ell m}^{\mathrm{(in)}}
}
       \label{greybodydefinition}
  \end{align}
under the condition that the radial wave is purely ingoing at the horizon.
This definition physically means,  in a 
time-reversed
sense, that $\Gamma$ is identical to the proportion that passes through the gravitational potential from the horizon toward the infinity without being reflected back.

In early 1970's, Teukolsky showed that the general wave equation in the four dimensional Kerr black hole background is separable into the angular and radial parts and developed analytic and numerical methods to solve the equation~\cite{Teukolsky:1973ha,Teukolsky2,Teukolsky3}.
Page then 
calculated the Hawking radiation spectra with the greybody factors taken into account~\cite{Page:1976df} and shown the whole time evolution of the four dimensional Kerr black hole~\cite{Page:1976ki}. 
The brane field equations that we have obtained in~\cite{Ida:2002ez} is a generalization of the Teukolsky equation.
In \cite{Ida:2002ez}, we have further obtained the analytic formulae of the greybody factors of the $s=0$, $1/2$ and $1$ brane fields in the low frequency limit $r_h\omega\ll1$ for $D=5$ dimensional $J\geq 0$ black hole.

Next goal is to obtain the greybody factors without relying on the low frequcy expansions. The numerical results with non-rotating $J=0$ black hole are shown by Harris and Kanti in~\cite{Harris:2003eg}. The greybody factors of $s=0$ brane field for general $J\geq 0$ hole
are presented by ourselves~\cite{Ida:2005zi,Ida:2005ax} and by Harris and Kanti~\cite{Harris:2004mf,Harris:2005jx}.
Recently there also appeared a paper that treats the $s=1$ modes for $J\geq0$ black hole in a restricted range $r_h\omega<4$~\cite{Casals:2005sa}.

In this paper, we complete the series of our work by presenting the greybody factors for $s=1/2$ and $s=1$ brane fields for the general $J\geq0$ hole
and the resultant Hawking spectra.
Now that all the brane field spectra are completed, we are finally able to show the whole time evolution of the black hole to confirm our previous claim that the spin-down phase, where the approximation $J=0$ is not valid, is hardly negligible: we find that typically much more than a half of the black hole mass is emitted in the spin-down phase.

The organization of the paper is as follows.
In section 2, we go beyond our previous treatment to obtain the brane field equations in a separable form for the lower spin components too
and recast the results into a rather simple formulae~\eqref{master-ang} and~\eqref{master-rad}.
We then show the resultant forms of the asymptotic solutions and the greybody factors.
In section 3, we explain our numerical methods to evade the contamination of the outgoing wave near the horizon when we impose the purely ingoing boundary condition there.
In section 4, we show our results of the greybody factors and the Hawking spectra for the brane-localized spinor and vector fields.
In section 5, we combine all the results of scalar~\cite{Ida:2005ax} and spinor/vector obtained in the previous section to determine the whole time evolution of the black hole.
The time evolution for the Randall-Sundrum (D=5) and the ADD (D=10) black hole are presented.

\section{Brane field equations}
The stationary rotating black holes would be described by
the Myers-Perry solution~\cite{Myers:1986un}
as indicated in~\cite{Gibbons:2002bh,Gibbons:2002av,Gibbons:2002ju,Morisawa:2004tc,Rogatko:2002bt,Rogatko:2003kj}.
The induced metric on the three-brane in the $(4+n)$-dimensional
Myers-Perry solution with a single nonzero angular momentum is given by
\begin{eqnarray}
g&=&{\Delta-a^2\sin^2\vartheta\over\Sigma}dt^2
+{2a(r^2+a^2-\Delta)\sin^2\vartheta\over\Sigma}dtd\varphi\nonumber\\
&&{}-{(r^2+a^2)^2-\Delta a^2\sin^2\vartheta\over\Sigma}\sin^2\vartheta d\varphi^2
-{\Sigma\over\Delta}dr^2-\Sigma d\vartheta^2,
\label{myers-perry}
\end{eqnarray}
where
\begin{align}
  \Sigma &= r^2+a^2\cos^2\vartheta, &
  \Delta &= r^2+a^2-\mu r^{1-n}.
  \end{align}
The parameters $\mu$ and $a$ are equivalent to the total mass $M$ and
the angular momentum $J$
\begin{align}
  M &= {(2+n)A_{2+n}\mu\over 16\pi G_{4+n}}, &
  J &= {A_{2+n}\mu a\over 8\pi G_{4+n}}
  \end{align}
evaluated at the spatial infinity of the $(4+n)$-dimensional space-time,
respectively,  where
$A_{2+n}=2\pi^{(3+n)/2}/\Gamma((3+n)/2)$ is the area of a unit $(2+n)$-sphere
and $G_{4+n}$ is the $(4+n)$-dimensional Newton constant of gravitation.

In Ref.~\cite{Ida:2002ez} it has been shown in terms of the Newman-Penrose formalism that
the spin $s=0$, $1/2$ and $1$ equations are of separable type in the
background space-time with the metric (\ref{myers-perry}).
Here we give the basic equations in more compactified notation.

The null tetrad $\{n,n',m,m'\}$ on the brane is
\begin{eqnarray}
n&=&\partial_t-a\sin^2\vartheta\partial_\varphi-{\Sigma\over \Delta}\partial_r,
\nonumber\\
n'&=&{\Delta\over 2\Sigma}(\partial_t-a\sin^2\vartheta\partial_\varphi)
+{1\over 2}\partial_r,
\nonumber\\
m&=&{i\sin\vartheta\over\sqrt{2}(r+ia\cos\vartheta)}\left[a\partial_t-(r^2+a^2)
\partial_\varphi\right]
-{r-ia\cos\vartheta\over\sqrt{2}}\partial_\vartheta,
\nonumber\\
m'&=&\bar m.
\end{eqnarray}
The spin-coefficients~\cite{Penrose:1985jw} in this tetrad system become
\begin{align}
  \tau  &= -{ia\sin\vartheta\over\sqrt{2}\Sigma}, &
  \rho  &= -{1\over r-ia\cos\vartheta}, &
  \beta &= -{\bar\rho\cot\vartheta\over 2\sqrt{2}}, \nonumber\\
  \tau' &= -{ia\rho^2\sin\vartheta\over\sqrt{2}}, &
  \rho' &= -{\rho^2\bar\rho\Delta\over 2}, &
  \epsilon'
        &= \rho'-{\rho\bar\rho\Delta_{,r}\over 4}, \nonumber\\
  \beta'&= \tau'+\bar\beta, &
  \kappa&= \sigma=\kappa'=\sigma'=\epsilon=0.
  \end{align}
The null tetrad basis vectors define the differential operators
\[
(D,D',m,m')=(\nabla_n,\nabla_{n'},\nabla_m,\nabla_{m'}).
\]
In terms of these differential operators and the spin-coefficients,
the field equations for $s=1/2$ fields $(\psi_0,\psi_1)$ (Weyl equation)
are given by
\begin{eqnarray}
D\psi_1-\delta'\psi_0&=&(\beta'-\tau')\psi_0+(\rho-\epsilon)\psi_1,
\label{NPWeyl-1}\\
\delta\psi_1-D'\psi_0&=&(\epsilon'-\rho')\psi_0+(\tau-\beta)\psi_1.
\label{NPWeyl-2}
\end{eqnarray}
On the other hand,
the field equations for $s=1$ fields $(\phi_0,\phi_1,\phi_2)$
(Maxwell equation) are given by
\begin{eqnarray}
D\phi_1-\delta'\phi_0&=&(2\beta'-\tau')\phi_0+2\rho\phi_1-\kappa\phi_2,
\label{NPMaxwell-1}\\
D\phi_2-\delta'\phi_1&=&\sigma'\phi_0-2\tau'\phi_1+(\rho-2\epsilon)\phi_2,
\label{NPMaxwell-2}\\
D'\phi_0-\delta\phi_1&=&(\rho'-2\epsilon')\phi_0-2\tau\phi_1+\sigma\phi_2,
\label{NPMaxwell-3}\\
D'\phi_1-\delta\phi_2&=&-\kappa'\phi_0+2\rho'\phi_1+(2\beta-\tau)\phi_2.
\label{NPMaxwell-4}
\end{eqnarray}
The all field variables are assumed to have time and angular dependence
$\phi_A,\psi_A\propto e^{-i\omega t+im\varphi}$.

Here it is useful to introduce the differential operators
\begin{align}
  \mathscr{D}
    &= \partial_r-i{K\over\Delta}, &
  \mathscr{D}^\dag
    &= \partial_r+i{K\over\Delta}, \nonumber\\
  \mathscr{L}_N
    &= \partial_\vartheta +Q+N\cot\vartheta, &
  \mathscr{L}^\dag_N
    &= \partial_\vartheta -Q+N\cot\vartheta \qquad (N=0,\pm {1/2},\pm 1)
  \end{align}
where
\begin{align}
  K(r)         &=  \omega(r^2+a^2)-ma, &
  Q(\vartheta) &= -\omega a\sin\vartheta+m\csc\vartheta
  \end{align}
have been defined.

\subsection{$s=1/2$ field equations}
The $s=1/2$ field equations (\ref{NPWeyl-1}) and (\ref{NPWeyl-2}) can be written as
\begin{eqnarray}
\mathscr{D}\Delta^{1\over 2}\eta_{-{1\over 2}}
&=&-{\mathscr L}_{{1\over 2}}\eta_{1\over 2}
\label{Unruh-1}\\
\mathscr{D}^\dag \Delta^{{1\over 2}}\eta_{1\over 2}
&=&{\mathscr L}^\dag_{1\over 2}\eta_{-{1\over 2}}
\label{Unruh-2}
\end{eqnarray}
where
\begin{eqnarray}
\eta_{-{1\over 2}}=\sqrt{2}\Delta^{-{1\over2}}\rho^{-1}\psi_1,~~
\eta_{{1\over 2}}=\psi_0
\end{eqnarray}
have been defined.
From
${\mathscr L}_{1\over 2}^\dag\times$ Eq.~(\ref{Unruh-1}) $+$
$\mathscr{D}\Delta^{1\over 2}\times$ Eq.~(\ref{Unruh-2})
we obtain
\begin{eqnarray}
\mathscr{D}\Delta^{1\over 2}
\mathscr{D}^\dag\Delta^{1\over 2} \eta_{1\over 2}
=-{\mathscr L}_{1\over 2}^\dag{\mathscr L}_{1\over 2}\eta_{1\over 2}.
\end{eqnarray}
By putting $\eta_{1\over 2}=R_{1\over 2}(r)S_{1\over 2}(\vartheta)e^{-i\omega t+im\varphi}$,
we have separated equations
\begin{eqnarray}
\mathscr{D}\Delta^{1\over 2}
\mathscr{D}^\dag\Delta^{1\over 2} R_{1\over 2}=\lambda_{{1\over 2}} R_{1\over 2},\\
{\mathscr L}_{1\over 2}^\dag{\mathscr L}_{1\over 2}S_{1\over 2}=-\lambda_{1\over 2}S_{1\over 2}
\label{ssp}
\end{eqnarray}
where $\lambda_{1\over 2}$ is the separation constant.
In a similar manner,
$\mathscr{D}^\dag \Delta^{1\over 2}\times$ (\ref{Unruh-1}) $-$
${\mathscr L}_{1\over 2}\times$ (\ref{Unruh-2})
and the substitution
$\eta_{-{1\over 2}}=R_{-{1\over 2}}(r)S_{-{1\over 2}}(\vartheta)e^{-i\omega t+im\varphi}$
give the separated equations
\begin{eqnarray}
\mathscr{D}^\dag\Delta^{1\over 2}
\mathscr{D}\Delta^{1\over 2}R_{-{1\over 2}}
&=&\lambda_{{1\over 2}}R_{-{1\over 2}},
\\
{\mathscr L}_{1\over 2}{\mathscr L}_{1\over 2}^\dag S_{-{1\over 2}}
&=&-\lambda_{{1\over 2}}S_{-{1\over 2}}.
\label{ssm}
\end{eqnarray}
The separation constant must again be $\lambda_{1\over 2}$, which can be
seen from explicit forms of angular equations~(\ref{ssp}) and (\ref{ssm}).
Hereafter, the angular function $S_A(\vartheta)$ is normalized such that
\begin{equation}
\int_0^\pi|S_A(\vartheta)|^2\sin\vartheta d\vartheta={1\over 2\pi}.
\end{equation}
Furthermore, from Eqs.~(\ref{Unruh-1}) and (\ref{Unruh-2}), we obtain
relationships
\begin{eqnarray}
S_{-{1\over 2}}&=&-{1\over\sqrt{\lambda_{{1\over 2}}}}{\mathscr L}_{{1\over 2}}S_{1\over 2}\\
S_{1\over 2}&=&{1\over \sqrt{\lambda_{1\over 2}}}{\mathscr L}_{1\over 2}^\dag S_{-{1\over 2}},
\end{eqnarray}
and
\begin{eqnarray}
\mathscr{D}\Delta^{1\over 2}R_{-{1\over 2}}&=&\sqrt{\lambda_{{1\over 2}}}R_{{1\over 2}}\\
\mathscr{D}^\dag\Delta^{1\over 2}R_{{1\over 2}}&=&
\sqrt{\lambda_{{1\over 2}}}R_{-{1\over 2}}
\end{eqnarray}
where the constant of proportionality can be determined by noting that
\begin{eqnarray}
\left({\mathscr L}_{1\over 2}S_{1\over 2}\right)^2\sin\vartheta
&=&\lambda_{1\over 2}S_{1\over 2}^2\sin\vartheta
+\partial_\vartheta\left(S_{1\over 2}{\mathscr L}_{1\over 2}S_{1\over 2}
\right)\\
\left({\mathscr L}_{1\over 2}^\dag S_{-{1\over 2}}\right)^2\sin\vartheta
&=&\lambda_{{1\over 2}}S_{-{1\over 2}}^2\sin\vartheta
+\partial_\vartheta\left(S_{-{1\over 2}}\sin\vartheta
{\mathscr L}_{1\over 2}^\dag S_{-{1\over 2}}\right)
\end{eqnarray}
hold.

\subsection{$s=1$ field equations}
The $s=1$ field equations
(\ref{NPMaxwell-1})-(\ref{NPMaxwell-4}) become
\begin{eqnarray}
\mathscr{D} \eta
&=&-\rho^{-2}{\mathscr L}_1\rho\eta_1,
\label{TMax-1}\\
{\mathscr L}_0\eta&=&
-\rho^{-2}\mathscr{D}\rho\Delta \eta_{-1},
\label{TMax-2}\\
{\mathscr L}_0^\dag \eta
&=&\rho^{-2}\mathscr{D}^\dag\rho\Delta\eta_1,
\label{TMax-3}\\
\mathscr{D}^\dag\eta&=&
\rho^{-2}{\mathscr L}_1^\dag\rho \eta_{-1},
\label{TMax-4}
\end{eqnarray}
where
\begin{eqnarray}
\eta_1=\phi_0,~~~\eta_{-1}=2\Delta^{-1}\rho^{-2}\phi_2,~~~\eta=\sqrt{2}\rho^{-2}\phi_1,
\end{eqnarray}
have been defined.
From
${\mathscr L}_0^\dag\times$ Eq.~(\ref{TMax-1}) $-$
$\mathscr{D}\times$ Eq.~(\ref{TMax-3}), we obtain
\begin{eqnarray}
&&\left[
(\mathscr{D}-\rho)(\mathscr{D}^\dag+\rho)\Delta
+({\mathscr L}_0^\dag-i \rho a\sin\vartheta)
({\mathscr L}_1+i \rho a\sin\vartheta)\right]\eta_1\nonumber\\
&&=\left(
\mathscr{D}\mathscr{D}^\dag\Delta
+{\mathscr L}_0^\dag{\mathscr L}_1
-{2i\omega\over\bar\rho}
\right)\eta_1=0.
\end{eqnarray}
Putting $\eta_1=R_1(r)S_1(\vartheta)e^{-i\omega t+im\varphi}$, we obtain separated
equations
\begin{eqnarray}
\left(\mathscr{D}\mathscr{D}^\dag\Delta
+2i\omega r\right)R_1=\lambda_1R_1
\\
\left({\mathscr L}_0^\dag{\mathscr L}_1
-2\omega a\cos\vartheta\right)S_1
=-\lambda_1S_1,
\end{eqnarray}
where $\lambda_1$ is the separation constant.
In a similar manner, from
$\mathscr{D}^\dag\times$ Eq.~(\ref{TMax-2}) $-$
${\mathscr L}_0\times$ Eq.~(\ref{TMax-4})
we obtain
\begin{eqnarray}
&&\left[(\mathscr{D}^\dag-\rho)(\mathscr{D}+\rho)\Delta
+({\mathscr L}_0-i\rho a\sin\vartheta)({\mathscr L}_1^\dag+i\rho a\sin\vartheta)
\right]\eta_{-1}\nonumber\\
&&=\left(\mathscr{D}^\dag\mathscr{D}\Delta+{\mathscr L}_0{\mathscr L}_1^\dag
+{2i\omega\over\bar\rho}\right)\eta_{-1}.
\end{eqnarray}
Putting $\eta_{-1}=R_{-1}(r)S_{-1}(\vartheta)e^{-i\omega t+im\varphi}$,
we obtain another set of separated equations
\begin{eqnarray}
\left(\mathscr{D}^\dag\mathscr{D}\Delta
-2i\omega r\right)R_{-1}=\lambda_{1}R_{-1}\\
\left({\mathscr L}_0{\mathscr L}_1^\dag+2\omega a\cos\vartheta\right)S_{-1}
=-\lambda_{1}S_{-1}
\end{eqnarray}
The constant of separation $\lambda_1$ is again common for both sets of equations.

The relationship between $\eta_1$ and $\eta_{-1}$ is seen as following.
From ${\mathscr L}_0\times$ Eq.~(\ref{TMax-1}) $-$
$\mathscr{D}\times$ Eq.~(\ref{TMax-2}), we obtain
\begin{eqnarray}
\mathscr{D}\mathscr{D}\Delta\eta_{-1}={\mathscr L}_0{\mathscr L}_1\eta_1.
\end{eqnarray}
From $\mathscr{D}^\dag\times$ Eq.~(\ref{TMax-3}) $-$
${\mathscr L}_0^\dag\times$ Eq.~(\ref{TMax-4}), we obtain
\begin{eqnarray}
\mathscr{D}^\dag\mathscr{D}^\dag\Delta\eta_1
={\mathscr L}_0^\dag{\mathscr L}_1^\dag\eta_{-1}.
\end{eqnarray}
These imply that
\begin{eqnarray}
{\mathscr L}_0{\mathscr L}_1 S_1&=& BS_{-1},\\
\mathscr{D}\mathscr{D}\Delta R_{-1}&=&BR_1,\\
{\mathscr L}_0^\dag{\mathscr L}_1^\dag S_{-1}&=&BS_1,\\
\mathscr{D}^\dag\mathscr{D}^\dag\Delta R_1&=&BR_{-1},
\end{eqnarray}
where $B$ is  a constant.
The constant $B$ is determined by
\begin{eqnarray}
B^2&=&\int ({\mathscr L}_0{\mathscr L}_1 S_{1})^2\sin\vartheta d\vartheta
=\int S_1{\mathscr L}_0^\dag{\mathscr L}_1^\dag {\mathscr L}_0{\mathscr L}_1 S_1
\nonumber\\
&=&\lambda_1^2-4\omega a(\omega a-m).
\end{eqnarray}
We choose
\begin{equation}
B=\sqrt{\lambda_1^2-4\omega a(\omega a-m)}.
\end{equation}

\subsection{Master equations for brane fields}
The angular and radial equations can be recasted into a neat form
\begin{eqnarray}
&&\Biggl[\left(\partial_\vartheta-{s\over|s|} Q+(1-|s|)\cot\vartheta\right)
\left(\partial_\vartheta+{s\over|s|} Q+|s|\cot\vartheta\right)
\nonumber\\
&&{}+(2|s|-1){({s\over|s|} Q\sin\vartheta)_{,\vartheta}\over\sin\vartheta}
+\lambda_{|s|}\Biggr]S_s=0,\\
&&\Biggl[\left(\partial_r-i{s\over|s|}{ K\over\Delta}\right)\Delta^{1-|s|}
\left(\partial_r+i{s\over|s|}{ K\over\Delta}\right)\Delta^{|s|}
\nonumber\\
&&{}+(2|s|-1)i{s\over|s|} K_{,r}-\lambda_{|s|}\Biggr]R_s=0,
\end{eqnarray}
More explicit forms are given by
\begin{eqnarray}
&&{1\over\sin\vartheta}{d\over d\vartheta}\left(\sin\vartheta{d S_s(\vartheta)\over d\vartheta}\right)
\nonumber\\
&&{}+\biggl[
(s-\omega a\cos\vartheta)^2-(s\cot\vartheta+m\csc\vartheta)^2
\nonumber\\
&&{}-\omega a(\omega a-2m)-|s|(|s|+1)+\lambda_{|s|}
\biggr]S_s(\vartheta)=0,
\label{master-ang} \\ \nonumber \\
&&\Delta^{-|s|}{d\over dr}\left(\Delta^{1+|s|}{dR_s(r)\over dr}\right)
\nonumber\\
&&{}+\left[
{K^2-isK\Delta_{,r}\over \Delta}+2i sK_{,r}+|s|\Delta_{,rr}-\lambda_{|s|}
\right]R_s(r)=0.
\label{master-rad}
\end{eqnarray}
These expressions are valid also for the massless scalar field by setting $s=0$.

Besides we have relationships between different radial components
\begin{eqnarray}
\left(\partial_r+i{K\over\Delta}\right) \Delta^{1\over 2}R_{1\over 2}
=\sqrt{\lambda_{1\over 2}}R_{-{1\over 2}}
\label{rel-spn}
\end{eqnarray}
for spinor fields and
\begin{eqnarray}
\left(\partial_r+i{K\over\Delta}\right)\left(\partial_r+i{K\over\Delta}\right)
\Delta R_1=\sqrt{\lambda_1^2-4\omega a(\omega a-m)}R_{-1}
\label{rel-vct}
\end{eqnarray}
for vector fields.
\subsection{Asymptotic solutions}
Asymptotic far field solution to the equation (\ref{master-rad}) is
\begin{eqnarray}
R_s\rightarrow
&&Y_s^{\mathrm{(in)}} {e^{-i\omega_*\xi_*}\over \xi^{1-s+|s|}}\left(1-i{\lambda_{|s|}+s-|s|
\over 2\omega\xi}\right)\nonumber\\
&&+Y_s^{\mathrm{(out)}} {e^{i\omega_*\xi_*}\over \xi^{1+s+|s|}}\left(1+i{\lambda_{|s|}-s-|s|
\over 2\omega\xi}\right),
\end{eqnarray}
where we have introduced the new radial coordinate
\begin{equation}
r_\star=\int^r{K\over\omega\Delta}
\end{equation}
and the dimensionless parameters
$\xi=r/r_h$, $\xi_*=r_\star/r_h$, $\omega_*=\omega r_h$ and $a_*=a/r_h$ have been
defined.

Then, Eqs.~(\ref{rel-spn}) and (\ref{rel-vct}) give the relationship between
coefficients
\begin{eqnarray}
Y_{-{1\over 2}}^{\mathrm{(out)}}&=&i{2\omega_*\over\sqrt{\lambda_{1\over 2}}}
Y_{1\over 2}^{\mathrm{(out)}},~~~
\label{inout-spn}
\\
Y_{-1}^{\mathrm{(out)}}&=&
-{4\omega_*^2\over\sqrt{\lambda_1^2-4\omega_* a_*(\omega_* a_*-m)}}Y_1^{\mathrm{(out)}}.
\label{inout-vct}
\end{eqnarray}

\subsection{Greybody factors}
The greybody factor is given by the absorption coefficient for
the wave equations.
By virtue of the Eqs.~(\ref{inout-spn}) and (\ref{inout-vct}),
the absorption coefficient can be calculated solely by solving
single radial equation.

The number flux vector for $s=1/2$ is
\begin{eqnarray}
j^\mu
=k\left(
n^\mu\psi_1\bar{\psi_1}
+n'{}^\mu\psi_0\bar{\psi_0}
-m^\mu\psi_1\bar{\psi_0}
-m'{}^\mu\psi_0\bar{\psi_1}
\right)
\end{eqnarray}
where $k$ is a constant.
The radial component become asymptotically
\begin{eqnarray}
j^r&=&k\left(|\psi_1|^2-{\Delta\over 2\Sigma}|\psi_0|^2\right)
=k{\Delta\over 2\Sigma}
\left(\bigl|\eta_{-{1\over 2}}\bigr|^2-\bigl|\eta_{1\over 2}\bigr|^2\right)
\nonumber\\
&&\rightarrow
{k\over 2\Sigma}\left(\bigl|Y_{-{1\over 2}}^{\mathrm{(out)}}\bigr|^2\bigl|S_{-{1\over 2}}\bigr|^2
-\bigl|Y_{1\over 2}^{\mathrm{(in)}}\bigr|^2\bigl|S_{1\over 2}\bigr|^2\right)
~~~(r\to +\infty).
\end{eqnarray}
Therefore, the absorption coefficient is given by
\begin{equation}
\Gamma_{1\over 2}=1-{
\dot{N_{1\over 2}^{\mathrm{(out)}}}\over \dot{N_{1\over 2}^{\mathrm{(in)}}}}
=1-{4\omega_*^2\over\lambda_{1\over 2}}
{|Y_{1\over 2}^{\mathrm{(out)}}\bigr|^2\over\bigl|Y_{1\over 2}^{\mathrm{(in)}}\bigr|^2},
\end{equation}
for spinor fields,
where $\dot N_{1\over 2}^{(\mathrm{in}/\mathrm{out})}$ denotes the total number flux.
The last expression is determined by solving only $s=1/2$ equation.

In a similar manner, we have the absorption coefficient
\begin{equation}
\Gamma_1=1-{\dot{
N_1^{\mathrm{(out)}}}\over \dot{N_1^{\mathrm{(in)}}}}
=1-{16\omega_*^4\over [\lambda_1^2-4\omega_* a_*(\omega_* a_*-m)]}
{|Y_1^{\mathrm{(out)}}\bigr|^2\over\bigl|Y_1^{\mathrm{(in)}}\bigr|^2},
\end{equation}
for vector fields.

\section{Numerical methods to obtain greybody factors}

We explain our numerical methods.
In this section, we take the unit
\begin{align}
  r_h&=1
  \end{align}
and always consider the case $s\geq0$ unless otherwise stated.
First, we switch from the Boyer-Lindquist frame~\eqref{myers-perry}
to the ingoing Kerr-Newman frame by:
\begin{align}
dv=&dt+{r^2+a^2\over\Delta}dr\\
d\widetilde \varphi=& d\varphi+{adr\over \Delta}
  \end{align}
The radial wave equation now becomes
\begin{align}
  \frac{d^2 R}{d r^2}+\eta \frac{d R}{d r} +\zeta R &= 0,
  \label{diffeqKN}
  \end{align}
with
\begin{align}
  {\eta} &= -\frac{(s-1)\Delta'+2iK}{\Delta}, &
  {\zeta} &=  \frac{2i \omega r(2s-1)-\lambda}{\Delta}
  \label{etatau}
  \end{align}
For the angular eigenvalue $\lambda$, we use the expansion of the form
\begin{align}
  \lambda = \sum_{n} C_n\,(a\omega)^n
  \end{align}
presented in \cite{Seidel:1988ue}. The coefficients $C_n$ are given for each angular mode $(l,m)$ and dumps quite fast, like a multiple of inversed factorials, as $l$ increases. We take up to $n=6$ terms. The accuracy of the approximation can be checked by the relative size of the last $n=6$ term to the sum when $a\omega$ is within the relevant region to the peak of the power spectrum, \textit{a posteriori}. We have checked that the ratio is at most of order .1\% for each relevant region.

The asymptotic solutions in the near horizon (NH) and far field (FF) regions are obtained as~\cite{Ida:2002ez}
\begin{align}
  R^{\mathrm{NH}} &\sim W_{\mathrm{in}}+W_{\mathrm{out}}e^{2ikr_*}\Delta^s, &
  R^{\mathrm{FF}} &\sim Y_{\mathrm{in}}r^{2s-1}+Y_{\mathrm{out}}e^{2 i k r_*}/r,
  \end{align}
where $r_*$ is the tortoise coordinate defined by $dr_*/dr=(r^2+a^2)/\Delta$ and $r_*\rightarrow r$ for $r\rightarrow\infty$.
Recall that $\Delta\rightarrow0$ for $r\rightarrow1$.
This coordinate change is not essential for the following analysis but convenient because the near horizon ingoing solution does not contain the tortoise coordinate $r_*$.

\subsection{Removing outgoing contamination at near-horizon}

For the case of scalar ($s=0$)~\cite{Ida:2005ax}, the calculation is simply to put the purely ingoing boundary condition $W_{\mathrm{in}}=0$ at a point $r_0=1+\epsilon$ close enough to the horizon $\epsilon\ll1$ and to solve the second order ordinary differential equation~\eqref{diffeqKN} numerically toward the far field region.
In this region we can easily read off the coefficients $Y_{\mathrm{in}},Y_{\mathrm{out}}$
by the $\chi^2$ fit, whose ratio directly leads to the greybody factor $\Gamma=1-|Y_{\mathrm{out}}/Y_{\mathrm{in}}|^2$.

Putting the purely ingoing boundary condition $W_{\mathrm{in}}=0$ at the near horizon $r=r_0$ is always polluted by a tiny outgoing wave, numerically. This itself is the case for the scalar too. The difficulty in the spinor ($s=1/2$) and vector ($s=1$) case is that the amplitude of the outgoing wave grows with respect to that of the ingoing one as we go apart from $r_0$.
To evade this problem, we first expand the near horizon solution
\begin{align}
  R^{\mathrm{NH}}_{\mathrm{in}}
    &= 1+ a_1 (r-1) + a_2(r-1)^2+ \cdots, \nonumber \\
  R^{\mathrm{NH}}_{\mathrm{out}}
    &= e^{2 i k r_*}(r-1)^s\left(1+ b_1(r-1) +\cdots \right),
       \label{NH_expansion}
\end{align}
where the coefficients $a_i,b_i$ are straightforwardly obtained by substituting the expansion of $\Delta,{\cal\eta},{\cal\zeta}$ and \eqref{NH_expansion} into \eqref{diffeqKN}, which serves linear equations for the coefficients.
The point to remove the outgoing contamination is the following subtraction
\begin{align}
  \widetilde{R} &= R-\left(1+a_1(r-1)\right),
  \end{align}
introduced in~\cite{Teukolsky2} as Bardeen's prescription.
Then $\widetilde{R}$ satisfies the equation
\begin{align}
  {\cal L} \widetilde{R} &= g,  \label{diffeq2}
  \end{align}
where
\begin{align}
  {\cal L}
    &= d^2/dr^2 +\eta\,d/dr +\zeta, \nonumber\\
  g &= -{\cal L}\,\left(1+ a_1 (r-r_H)\right)
     = -\eta a_1-\zeta\left(1+ a_1 (r-r_H)\right).
  \end{align}
Now \eqref{diffeq2} can be safely solved toward the far field region without the growing outgoing contamination.

\subsection{Matching at far field}
We expand the far field solutions for spinor $(s=1/2)$
\begin{align}
  R^{\mathrm{FF}}_{\mathrm{in}}
    &= 1+ \frac{c^f_1}{r} + \frac{c^f_2}{r^2}+\frac{c^f_3}{r^3} +\cdots , \nonumber \\
  R^{\mathrm{FF}}_{\mathrm{out}}
    &= e^{2 i k r_*}\frac{1}{r} \left( 1+
\frac{d^f_1}{r}+\frac{d^f_2}{r^2}\cdots\right),
  \label{FF_expansion}
  \end{align}
and for vector ($s=1$)
\begin{align}
  R^{\mathrm{FF}}_{\mathrm{in}}
    &= r \left(1+\frac{c^v_1}{r}
                +\frac{c^v_2}{r^2}
                +\frac{c^v_3}{r^3}+\cdots\right), \nonumber\\
  R^{\mathrm{FF}}_{\mathrm{out}}
    &= e^{2 i k r_*}\frac{1}{r}
       \left(1+\frac{d^v_1}{r}
              +\frac{d^v_2}{r^2}
              +\frac{d^v_3}{r^3} +\cdots\right).
  \end{align}
Again the coefficients $c_i^{f,v},d_i^{f,v}$ can be straightforwardly obtained.
The asymptotic form of the $\widetilde{R}_s$ is now
\begin{align}
  \widetilde{R}_{1/2}
    &= \left(Y_{\mathrm{in}}+1-a_1\right)
       +a_1 r
       +Y_{\mathrm{in}}\frac{c^f_1}{r}
       +Y_{\mathrm{out}}\frac{e^{2i\omega r_*}}{r}, \\
  \widetilde{R}_1
    &=  \left(Y_{\mathrm{in}}-a_1\right)r
       +\left(Y_{\mathrm{in}}c^v_1-1+a_1\right)
       +Y_{\mathrm{in}}\frac{c^v_2}{r}
       +Y_{\mathrm{out}}\frac{e^{2i\omega r_*}}{r}.
  \end{align}
By the $\chi^2$ fit to the obtained solution $\widetilde{R}$ in the far field region, we can determine $Y_{\mathrm{in}},Y_{\mathrm{out}}$
since the bigger terms than that contains $Y_{\mathrm{out}}$ are all fixed.
The smallness of the oscillating term containing $Y_{\mathrm{out}}$
can be easily overcome by keeping sufficient digits in the numerics and by taking sufficiently dense sample points in the far field region for the $\chi^2$ fit.

Finally the greybody factors can be obtained as
\begin{eqnarray}
  \Gamma_{s=1/2}
    &= 1-\frac{2\omega}{\left|c^f_1\right|}
         \left|{Y_{\mathrm{out}}\over Y_{\mathrm{in}}}\right|^2, \\
  \Gamma_{s=1}
    &= 1-\frac{2\omega^2}{\left|c^v_2\right|}
         \left|{Y_{\mathrm{out}}\over Y_{\mathrm{in}}}\right|^2.
\end{eqnarray}
%

\section{Greybody factors and Hawking radiation spectra}

In~\cite{Ida:2002ez} we have argued that the black hole production cross section for a center of mass energy $s$ is well approximated by
\begin{align}
  {d\sigma\over dJ}(s)
    =&
    \begin{array}{ll}
      8\pi J/s & \mathrm{for} J<J_{\mathrm{max}},\\
         0     & \mathrm{for} J>J_{\mathrm{max}},
    \end{array}
  \end{align}
except for the region where $J$ is very close to $J_{\mathrm{max}}$, and consequently that the black hole tends to be produced with large angular momentum.
The rotation parameter $a$ corresponding to $J_{\mathrm{max}}$ is~\cite{Ida:2002ez}
\begin{align}
  a_{\mathrm{max}} &= {n\over2}+1
  \label{amax}
  \end{align}
for $D=4+n$ dimensional black hole.

Once produced, we assume that the decay process of the black hole is governed by the Hawking radiation into the brane-localized standard model fields
See~\cite{Cardoso:2005jq} for a 
review on the estimation of the amount of energy radiated at the black hole formation process (balding phase) rather than by the Hawking radiation and see also~\cite{Yoshino:2005ps} for the recent progress. 
in the spirit that the quantum gravitational correction will be read off as a deviation from this precise prediction in the black hole picture.

With this assumption in mind,
the number of spin $s$ particles emitted into an $(\ell,m)$ angular mode
of the spheroidal harmonics is given, per degree of freedom per unit time per energy interval $[\omega,\omega+d\omega]$, by
\begin{align}
  {dN_{s,l,m}\over d\omega dt}
    &= {1\over2\pi}\frac{\Gamma_{s,l,m}(\omega)}{e^{(\omega-m\Omega)/T}-(-1)^{2s}},
       \label{number_spectrum}
  \end{align}
where $T=[(n+1)+(n-1)a_*^2]/4\pi r_h(1+a_*^2)$ and $\Omega=a_*/(1+a_*^2)r_h$ are the Hawking temperature
and angular velocity of the black hole, respectively.
The corresponding power and angular spectra are obtained by multiplying the number spectrum~\eqref{number_spectrum} by $\omega$ and $m$, respectively.

The time evolution of the black hole with mass $M$ and angular momentum $J$ is then governed by
\begin{eqnarray}
-\frac{d}{dt}\left(%
\begin{array}{c}
  M \\
  J \\
\end{array}%
\right) = \sum_{s,l,m}g_s \int_0^\infty  d\omega {dN_{s,l,m}\over d\omega dt}\left(%
\begin{array}{c}
  \omega \\
  m \\
\end{array}%
\right), \label{rates}
\end{eqnarray}
where $g_s$ is the number of massles degrees of freedom at
temperature $T$, namely, the number of degrees of freedom whose
masses are smaller than $T$ with spin $s$.

\subsection{Greybody factors for spinor and vector fields}
We present the greybody factors obtained by the procedure above.
Hereafter we limit ourselves to the case $D\leq 11$ motivated by the fact that $D=11$ is the highest possible dimension if we exclude the $s>2$ component fields from the supergravity multiplet, though in principle we can consider a larger dimensional case as well~\cite{Hewett:2005iw}.

In 
Fig.~1, 
we plot the greybody factors for the brane-localized spinor field for the non-rotating ($a_*=0$) and highly-rotating ($a_*=1.5$) black holes. Note that $a_*=1.5$ is the highest possible rotation parameter for a $D=5$ black hole when the black hole is produced by a collision of two particles, see Eq.~\eqref{amax}.

The Figure~2
shows the corresponding plots for the vector field ($s=1$).
We observe that the greybody factor becomes negative when $\omega<m\Omega$, rendering the number spectrum~\eqref{number_spectrum} always positive.
In other words, when the black hole is highly rotating, an incoming steady energy flux with $\omega<m\Omega$ is scattered back by the hole with an increased amplitude, 
see Eq.~\eqref{greybodydefinition}. This is called the superradiance~\cite{PressTeukolsky}.
Though we cannot observe the superradiance itself for the TeV black hole since it is decaying so fast, we show the rate of the energy amplification in
 Fig.~3
for its own interest.

\subsection{Power and angular spectra for spinor}
Hereafter in this section, we present the dimensionless power and angular spectra per degree of freedom:
\begin{align}
  r_h{dE_{s,l,m}\over d\omega dt}
    &= {1\over2\pi}\frac{r_h\omega\,\Gamma_{s,l,m}(\omega)}{e^{(\omega-m\Omega)/T}-(-1)^{2s}}, &
  {dJ_{s,l,m}\over d\omega dt}
    &= {1\over2\pi}\frac{m\,\Gamma_{s,l,m}(\omega)}{e^{(\omega-m\Omega)/T}-(-1)^{2s}},
  \end{align}
versus the dimensionless energy of the emitted particle $r_h\omega$.
Note that this implies that the frequency $\omega$ in the horizontal axis is given in unit of $r_h^{-1}$, which varies for a fixed mass $M$ when we vary the angular momentum $J$.
For simpler presentation we show our plot for $0<r_h\omega<8$ in this section but
we have also calculated all the $\omega$ regions for $a_*\leq 1.5$ to obtain the total power and angular spectra
in the next section, see also 
Figs.~14.

In 
Figs.~4 and 5,
we plot them for the brane-localized spinor and vector fields, respectively,
with varying black hole rotation $a_*$ for fixed $D=5$ (Randall-Sundrum black hole) and $D=11$ (ADD black hole).
It is clear that the larger the rotation parameter $a_*$ is, the more enhanced both the power and angular spectra are.

In 
Figs.~6 and 7, 
we plot the same
for spinor and vector, respectively, for varying dimensions $D$ with fixed black hole rotation $a_*=0$ (non-rotating) and $a_*=1.5$ (highly-rotating).
These figures show that the black hole radiation is greatly enhanced for larger dimensions.
Note that the peaks coming from $e=m$ modes are distinctive
only for the Randall-Sundrum black hole,
especially in the case of the spinorial emission.

The substructure behind the total spectrum is shown in 
Figs.~8,9,10
and 
in Figs.~11,12,13.
For each column, the greybody factor, power spectrum,
and the angular spectrum are shown from above to below.
We can see that each region of the greybody factor's rise in $r_h\omega$ coincides
with each peak.
For non-rotating hole, the angular modes $m=-l,-l+1,\dots,l$ give exactly the same greybody factors and the $\pm m$ modes cancel each other in each given $l$ mode for power and angular spectra.
When the black hole is highly rotating, the contributions from the $l=m$ modes become dominant.

For a highly rotating black hole with large $a_*$,
the contributions from $l=m$ modes become dominant,
which greatly enhance the amplitude of total spectra at high frequency
region $r_h\omega>1$.
This is sometimes called the superradiant enhancement of the higher spin emission, though it is not directly related to the original meaning~\cite{PressTeukolsky} of 
the superradiance as shown in 
Fig.~3
To exhibit such a high frequency tails which are taken into account in the calculation of the next section, we present 
Figure~14.


\section{Time evolution of black hole}
In Fig.~15,
we show the schematic pictures for the black hole life at various stages: production phase, the balding phase, the spin-down phase, the Schwarz\-schild phase, and the Planck phase~\cite{Giddings:2001bu}. The spin-down and Schwarz\-schild phases are of interest here.

We show the black hole time evolution as it emitts brane
localized particles.
We calculate the rates at which
energy and angular momentum are radiated into the brane-localized
standard model fields and the evolution of the mass
and the angular momentum.
A particular quantity of interest is the portion of the energy emitted during the spin-down phase and the (almost) Schwarz\-schild phase.

\subsection{Setup for time evolution}
To follow the time evolution, it is convenient to make quantities
invariant under the scaling of $r_s=(1+a_*^2)^{1/(n+1)}r_h$,
which does not vary for a fixed $M$ when we vary $J$.
We define a scale invariant function
$\gamma(a_s)$, with respect to  the scale invariant rotation
parameter $a_s= a/r_s$, as follows:
\begin{eqnarray}
\gamma^{-1}(a_s)&\equiv& \frac{d \ln a_s}{d \ln
M/M_i}\label{h-function}\\
&=&\frac{n+2}{2}\left(\frac{1}{a}\frac{d J}{ d
M}-\frac{2}{n+1}\right),
\end{eqnarray}
where the mass of a hole is measured in the unit of the initial mass
$M_i$. This quantity is directly integrated. We calculate the
ratio of the final mass $M_f$ to the initial mass $M_i$
by integrating Eq.~\eqref{h-function} with an initial
rotation parameter $a_s(\mathrm{ini})$
\begin{eqnarray}
\frac{M_f}{M_i}=\exp\left(\int_{a_s(\mathrm{ini})}^{a_s(\mathrm{final})}
d a_s \frac{\gamma(a_s)}{a_s }\right).
\end{eqnarray}
The amount of energy which is radiated in spin-down phase ($0
\thickapprox a_s(\mathrm{final}) \leqslant a_s \leqslant
a_s (\mathrm{ini})$) is $(M_i-M_f)$ and then the remaining $M_f$ will be subsequently
radiated off in the Schwarz\-schild phase,
where the angular momentum of black hole is vanishing.

We consider the evolution of the black hole.
Since the time roughly scales as $r_s^{n+3}$ in $(4+n)$ dimensions,
it is convenient to introduce scale invariant rates for energy and
angular momentum as follows.\footnote{We can 
understand this by simply looking at the formula $-dM/dt \sim A T^4$
where the surface area of horizon $A\sim r_s^2$ for brane fields and
the temperature of the hole $T\sim 1/r_s$ and $M\sim r_s^{n+1}$.}
\begin{eqnarray}
\alpha (a_s) &\equiv& - r_s^{n+3}\frac{d \ln M}{d t},\\
\beta (a_s) &\equiv& - r_s^{n+3}\frac{d\ln J}{d t},
\end{eqnarray}
with these new variables $\gamma(a_s)$ can be written as
$\gamma^{-1}(a_s)= \beta/\alpha(a_s) -(n+2)/(n+1)$. For all the
standard model particles,
\begin{eqnarray}
\alpha_{\mathrm{SM}} &=& g_s \alpha_{s=0} + g_f \alpha_{s=1/2}+ g_v
\alpha_{s=1}, \\
\beta_{\mathrm{SM}} &=& g_s \beta_{s=0} + g_f \beta_{s=1/2}+ g_v
\beta_{s=1},
\end{eqnarray}
where $g_s=4$, $g_f=90$ and $g_v=24$ are adopted in this section.

We also introduce dimensionless variables $y$ and $z$ to take
angular momentum and mass of the hole:
\begin{eqnarray}
y &\equiv& -\ln a_s, \\
z&\equiv& -\ln \frac{M}{M_i},
\end{eqnarray}
then finally we get the time variation of energy and angular
momentum in terms of scale-invariant time parameter
($\tau=r_s^{-n-3}(\mathrm{ini})t$) with initial mass of the hole:
\begin{eqnarray}
\frac{d z}{d y }&=& \frac{\alpha}{\beta-\alpha \left(\frac{n+2}{n+1}\right)}, \nonumber \\
\frac{d y}{d \tau}&=& (\beta- \alpha\left(\frac{n+2}{n+1}\right))
~e^{\frac{n+3}{n+1}z}. \label{DE for z,y}
\end{eqnarray}

After finding the solutions $z(y)$ and $\tau(y)$ of the coupled
differential equations \ref{DE for z,y}, one can get $y(\tau)$ and
$z(\tau)$, hence $a_s$ and $M/M_i$, as a function of time. From
these, one can find how other quantities evolve, such as the area.

For our purpose, it is convenient to convert the set of variables $(r_h,a_*)$
For conversion of unit, the following expressions
are useful with $a_s = a_* /(1+a_*^2)^{1/(n+1)}$.
\begin{eqnarray}
\alpha(a_s)&=&- \iota_n^{n+1}(1+a_*^2)^{\frac{2}{n+1}}r_h^2 \frac{d M}{dt},\\
\beta(a_s)&=&-\kappa_n^{n+1}(1+a_*^2)^{\frac{2}{n+1}}r_h^2
\frac{1}{a}\frac{d J}{dt},
\end{eqnarray}
where
\begin{eqnarray}
\iota_n &=& r_s M^{-\frac{1}{n+1}}=\left(\frac{16\pi G}{(n+2)\Omega_{n+2}}\right)^{\frac{1}{n+1}},\\
\kappa_n &=& \iota_n (\frac{n+2}{2})^{\frac{1}{n+1}}.
\end{eqnarray}

\subsection{Results for time evolution}
In this section we use the natural unit $8 \pi G=1$.
We assume that all the standard model fields are massless
and therefore the effective degrees of freedom
are given by $g_s=4$, $g_f=90$ and $g_v=24$.
%
In Fig.~16,
we draw the rates at which the
energy and angular momentum are radiated into the brane localized
standard model fields.
We have explicitly calculated the rates up to the rotation parameter $a_* < 1.5$ for the Randall-Sundrum black hole ($D=5$) and for the ADD black hole ($D=10$).
For $D=10$ we have extrapolated our result using the cubic-curve approximation up to $a_*<4$ to cover all the possible rotation, see Eq.~\eqref{amax}.
The curves are plotted with respect to $a_s$ rather than $a_*$ using the conversion rules described in the previous section, namely up to
$a_s<0.83$ and 2.67 for $D=5$ and 10, respectively.
The $D=10$ results are exact without using the cubic-curve extrapolation up to $a_s<1.27$.

In Fig.~16,
one can see that the power $\alpha$ and torque $\beta$ are increasing functions of the angular momentum $a_s$.
The vector emission dominates over the spinor and scalar emission for the high rotation parameter, sometimes called the superradiant enhancement of the higher spin particle emission, but as rotation becomes slower the 
fermion channel becomes increasingly important.
Generally, angular momentum is emitted much faster than energy, therefore rapidly rotating black hole spins down to a nearly non-rotating state before its mass has been 
radiated off completely.

In Fig.~17,
we plot the mass of a hole as 
a function of the rotation $a_s$
for the virtual setup where only the scalar (s),
spinor (f), or vector (v) field is emitted, respectively,
and for the realistic case where all the standard model fields are
emitted (SM).
One can see that the larger the particle's spin is, the more effectively the black hole angular momentum is carried away.
For the most effective case of vector-only (v), the angular momentum is carried away so rapidly that more than 30\% of the mass still remains after the spin-down phase, to be radiated off at the Schwarz\-schild phase.
In contrast, the scalar-only case (s) exhibits that the black hole radiates its whole mass before it stops rotation.
For the realistic case (SM), black hole loses roughly $70\%$ to $80\%$ in $D=5$ and $D=10$, respectively, before it stops rotation when starting from the maximum rotation.
Note that the scalar emission (s) is subdominant comparing to the vector (v) and spinors (f) because of its small effective degrees of freedom and small emission rates.

In Figs.~18 and 19,
we plot the time evolution of rotation
parameter and mass 
The unit time $t_0$ is defined by the time duration
from the initial state to the state with $a_s=0.01$
for the virtual cases of spinor-only (f) and vector-only (v).
For scalar-only case (s), the mass goes to zero before rotation stops,
therefore we defined $t_0$ for scalar to be the duration until
the whole mass is radiated off.
The initial rotation parameter is fixed by $a_s=0.83$ and $2.67$
in $D=5$ and $D=10$, respectively. The mass of the hole goes
to zero before the rotation parameter goes
to zero when only scalar emission is available.
We have taken the initial radiation parameter to be $a_s=0.83$ and $2.67$
for $D=5$ and 10 that are maximally rotations allowed by the initial collision.

%
When all the standard model fields are turned on (SM), the evolution is essentially determined by the spinor and vector radiation.
The figures show that a black hole spins down quickly at the first stage with large rotation parameter and the decrease of rotation parameter slows down as angular momentum of the hole is reduced.

\section{Summary and Discussion}
We have generalized our previous result~\cite{Ida:2002ez}
for the greybody factor and Hawking radiation spectrum
of the higher dimensional Kerr black hole
to general dimensions $D=4+n$, 
without relying on the low frequency limit $r_h\omega\ll1$,
not only for the scalar field ($s=0$)~\cite{Ida:2005ax}
but also for the spinor ($s=1/2$) and vector fields ($s=1$).
Now we can completely describe the evolution of a
black hole with any given initial mass and angular momentum by
taking into account any type of field residing on the brane.

We have developed the numerical method to solve the radial
Teukolsky equation which has been generalized to the higher dimension
($D=4+n$) in the first paper~\cite{Ida:2002ez} of our series of publication.
There are two points in our numerical methods.
First, we have imposed the proper purely-ingoing boundary condition
near the horizon without the growing contamination of the out-going wave
by extracting lower order terms explicitly.
Second, we have developed the method to 
fit the in-going and out-going part from the numerically
integrated wave solution at far field region
by explicitly obtaining the next-to-next order expansion
(or next-to-next-to-next order in vector case) of the solution.
With these progress in numerical treatment, we can
safely integrate the generalized Teukolsky equation
up to very large $r$ without out-going wave contamination.

Then we have calculated all the possible modes
to completely determine
the radiation rate of the mass and angular momentum of the hole.
Totally 3407 modes
are computed explicitly, other than the modes which
are confirmed to be negligible.
A black hole tends to lose its angular momentum
at the early stage of evolution.
However the black hole still have a sizable rotating
parameter after radiating half of its mass.
Typically more than $70\%$ or $80\%$ of black hole's mass is lost
during the spin down phase.
In the case of very fast initial rotation, the number could be modified quantitatively by taking the bulk graviton emission into account, especially for a larger number of extra dimensions as discussed in the Introduction, but the result would remain the case qualitatively.\footnote{We note that a friction between the brane and black hole can be sizable and can also reduce the angular momentum faster at the initial stage~\cite{Frolov:2004wy,Frolov:2004bq}.}

We have determined the radiation and evolution
of the spin-down and Schwarz\-schild phases up to the ambiguities for the initial fastest rotations shown above. The remaining hurdle
is the evaluation of the balding phase, which is still being disputed
due to its non-purturbative nature,
to extract the quantum gravitational information at the Planck phase from the experimental
data at LHC.

\vspace{.5cm}
\noindent
{\bf Acknowledgement}\\
We would like to thank
K.\ Hagiwara, 
Y.\ Okada, 
V.\ S.\ Rychkov, 
S.\ A.\ Teukolsky, and
H.\ Yoshino
for useful discussion.
K.O.\ is supported by EU grant MRTN-CT-2004-503369,
and
S.C.P.\ is supported by the Korea Research Foundation Grant funded
by the Korean Government (MOEHRD), grant KRF-2005-214-C00147.

\bibliography{IOP}
\bibliographystyle{utphys}

\begin{figure}[tbp]
\begin{center}
  \includegraphics{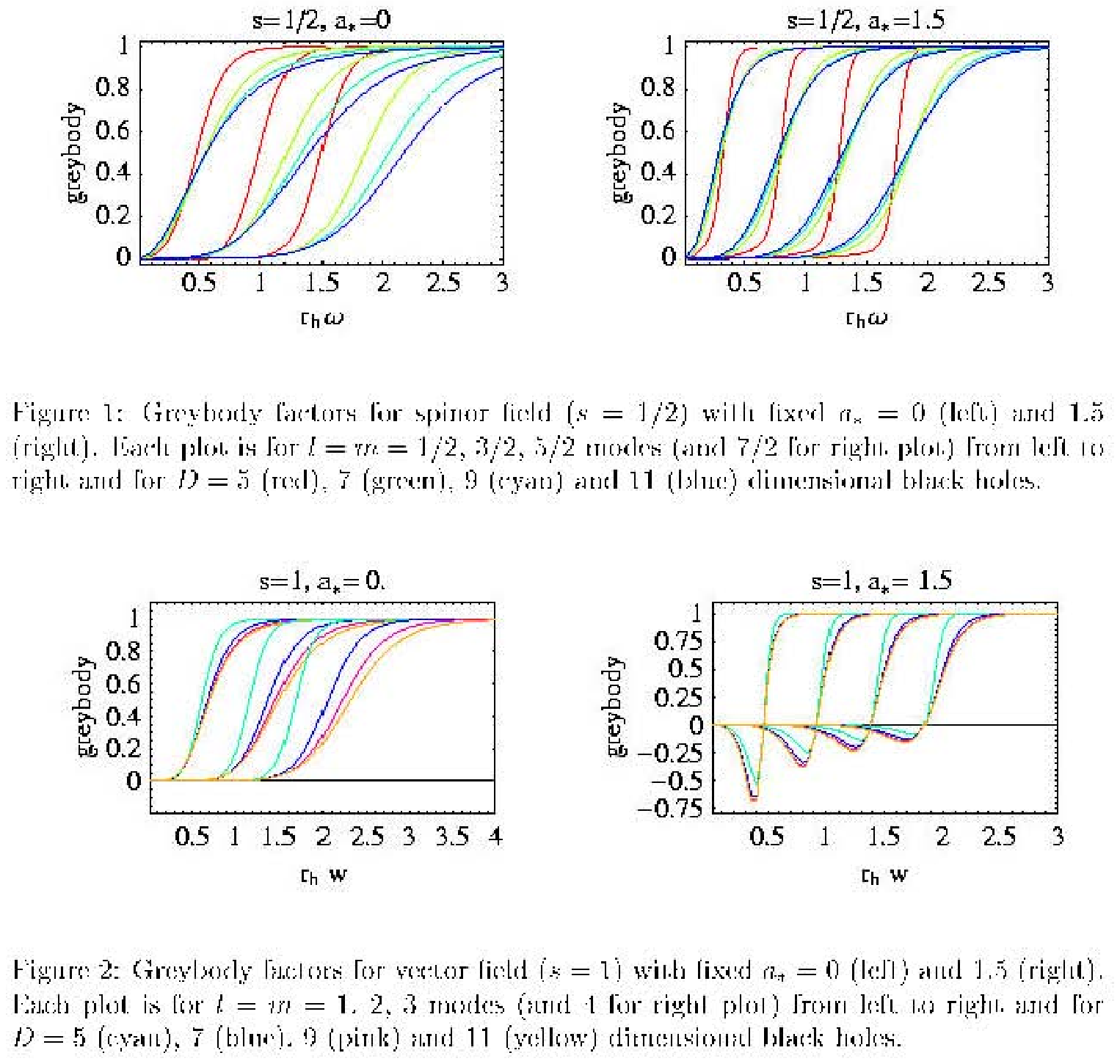}
\end{center}
\end{figure}

\begin{figure}[tbp]
\begin{center}
  \includegraphics{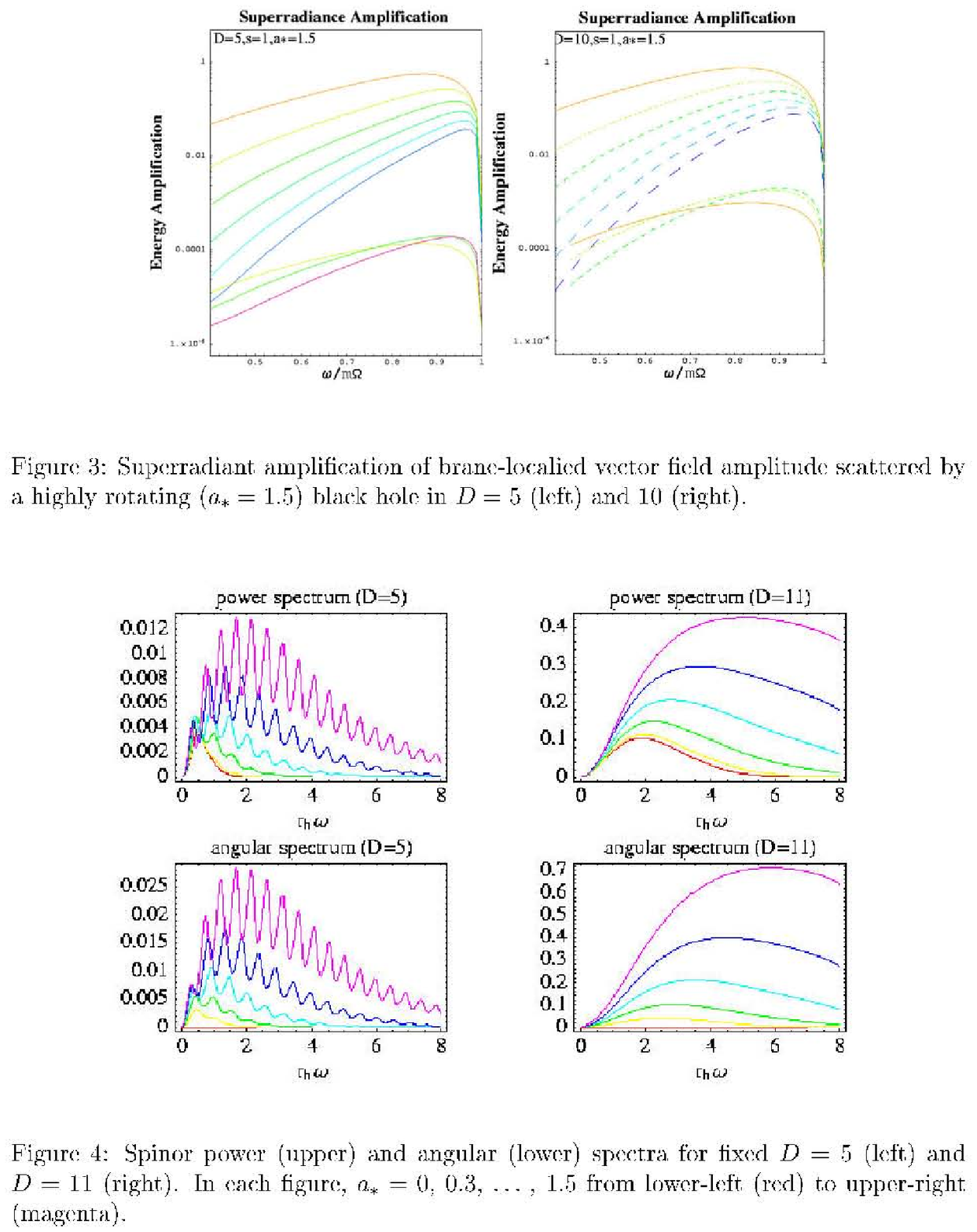}
\end{center}
\end{figure}

\begin{figure}[tbp]
\begin{center}
  \includegraphics{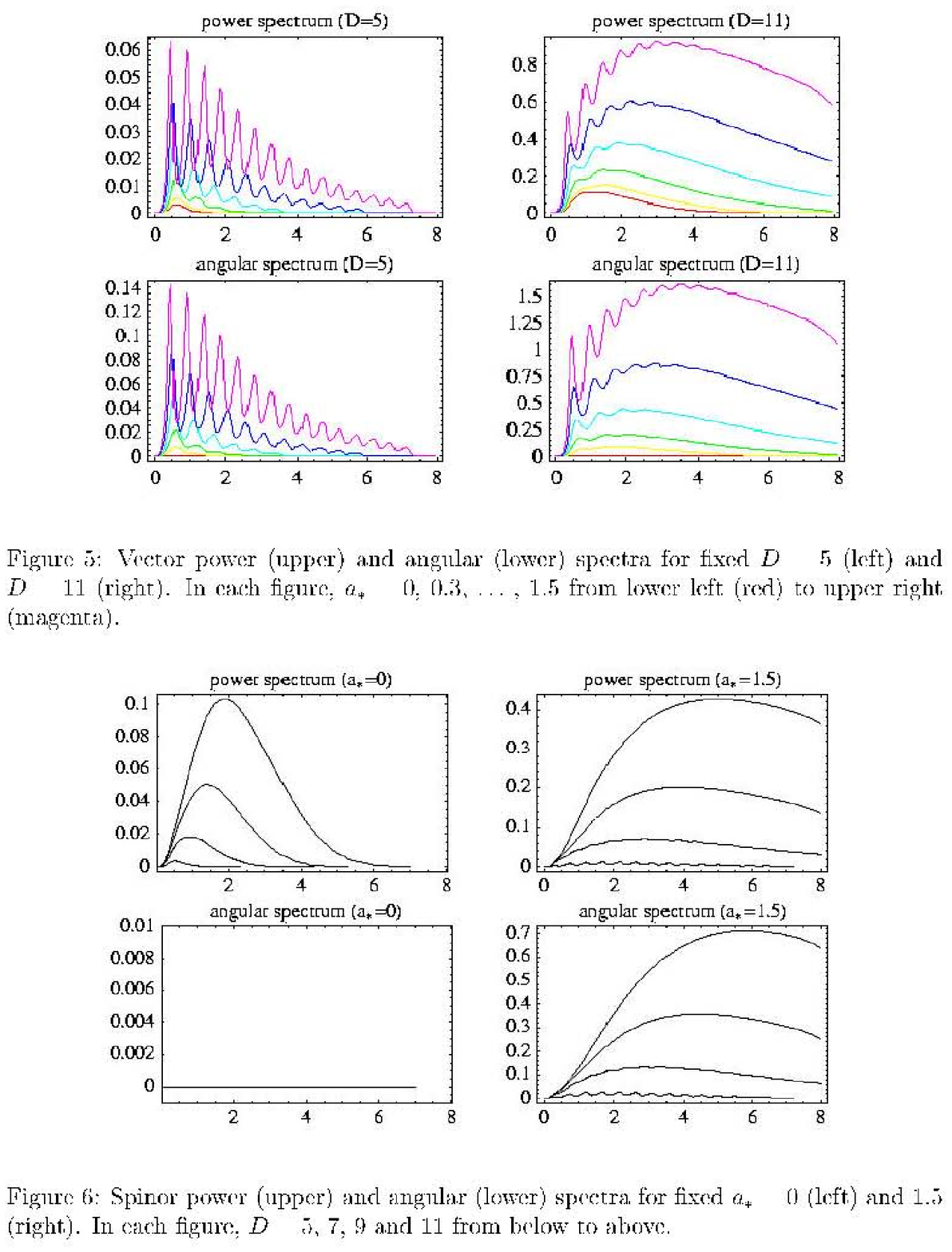}
\end{center}
\end{figure}

\begin{figure}[tbp]
\begin{center}
  \includegraphics{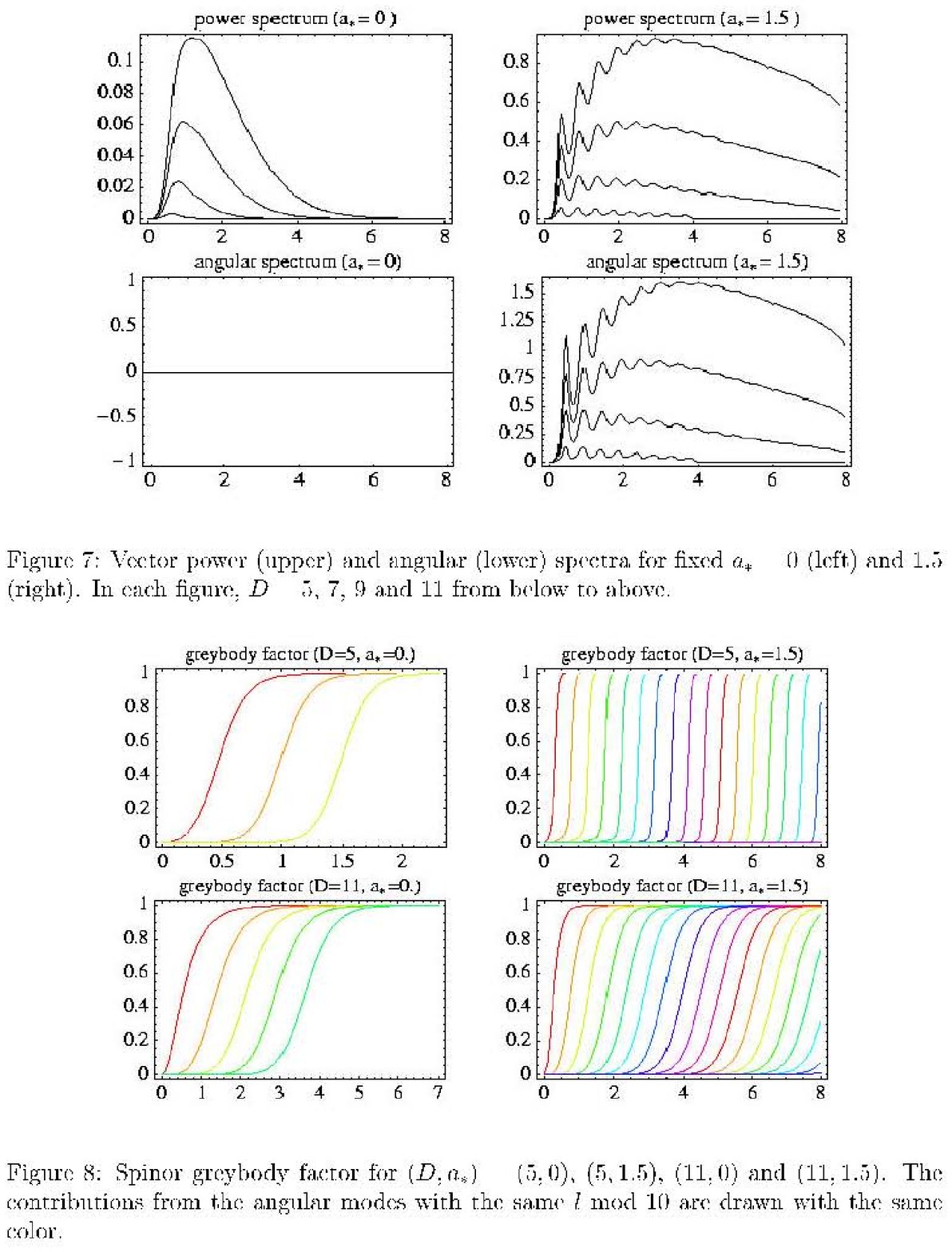}
\end{center}
\end{figure}

\newpage
\begin{figure}[tbp]
\begin{center}
  \includegraphics{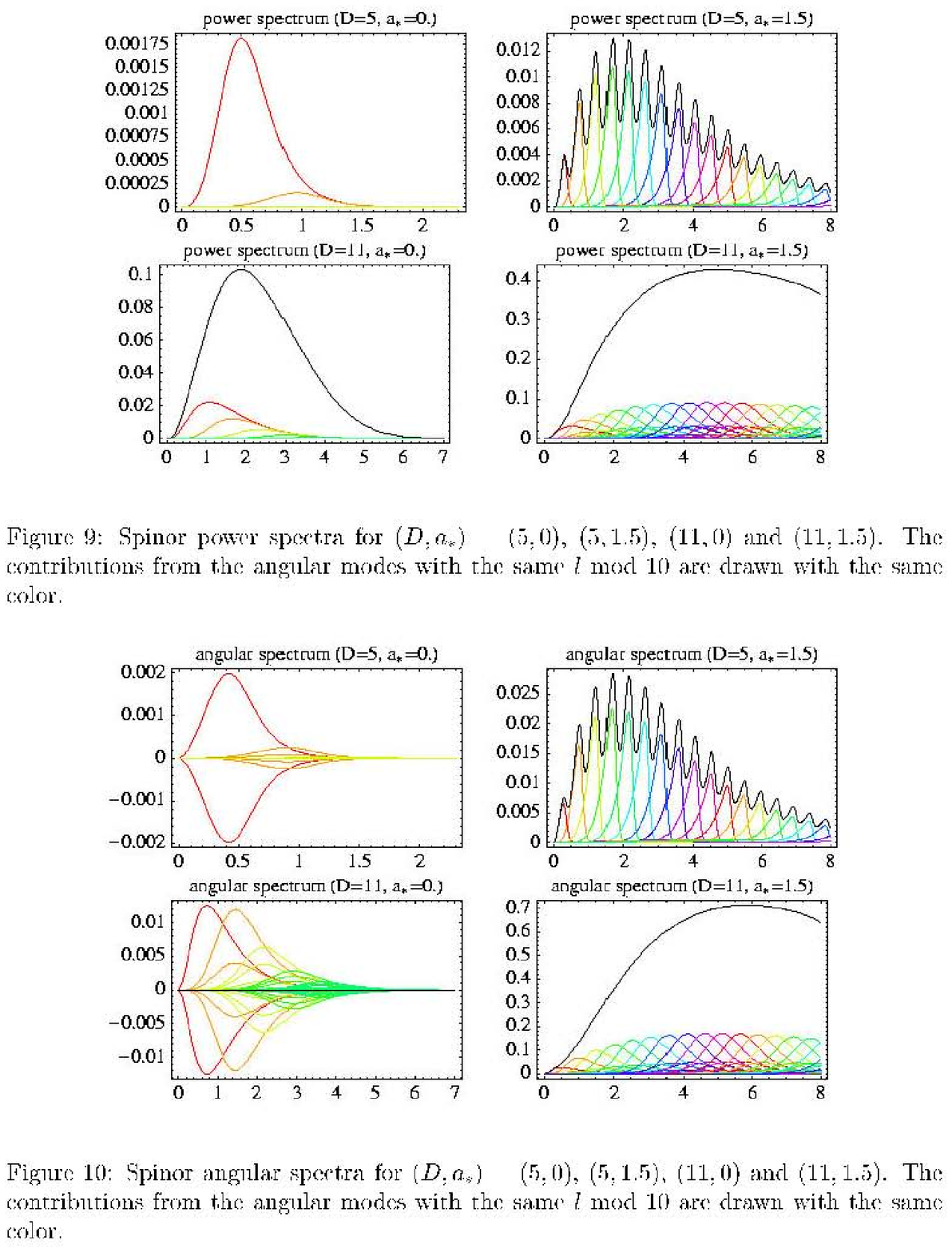}
\end{center}
\end{figure}

\begin{figure}[tbp]
\begin{center}
  \includegraphics{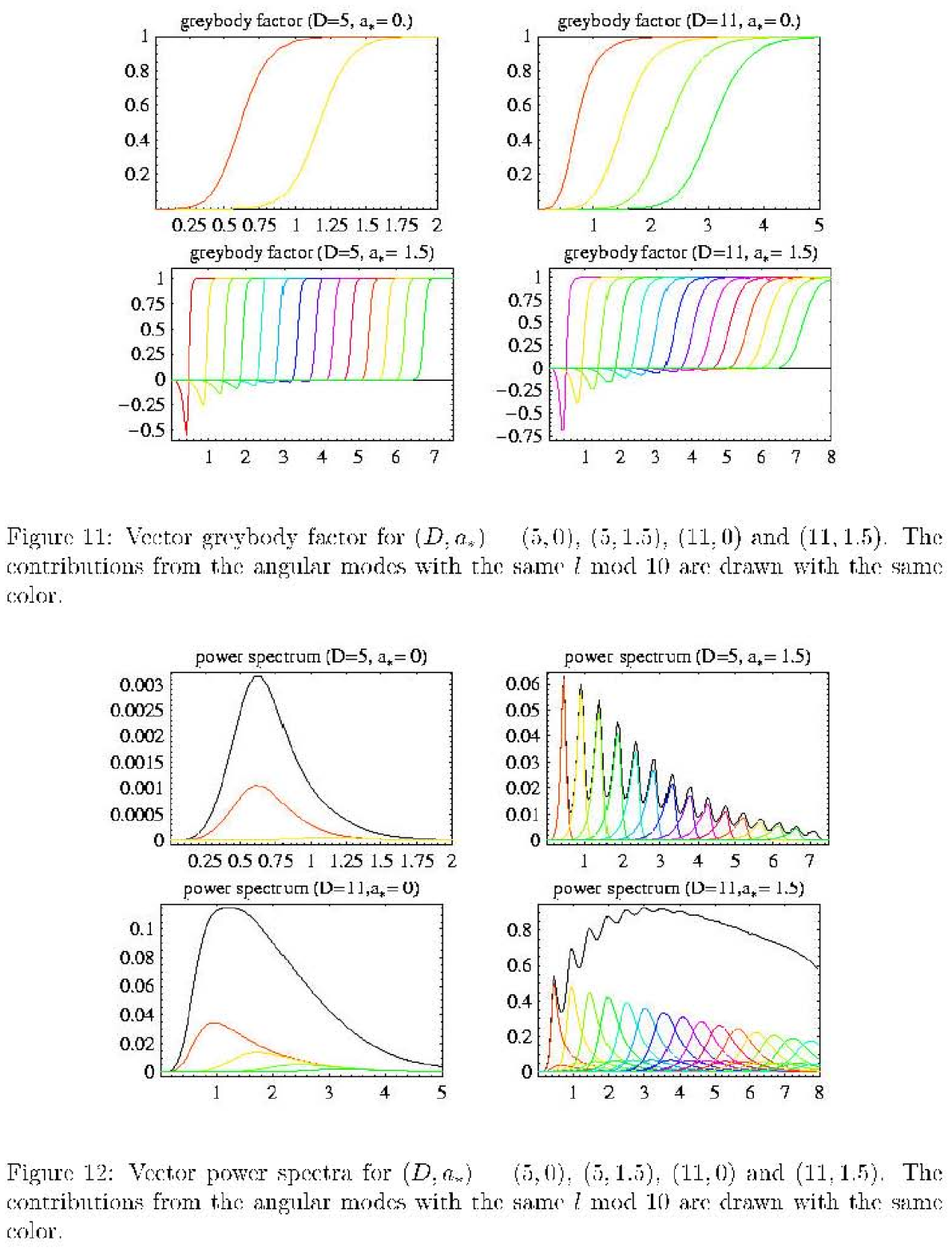}
\end{center}
\end{figure}

\begin{figure}[tbp]
\begin{center}
  \includegraphics{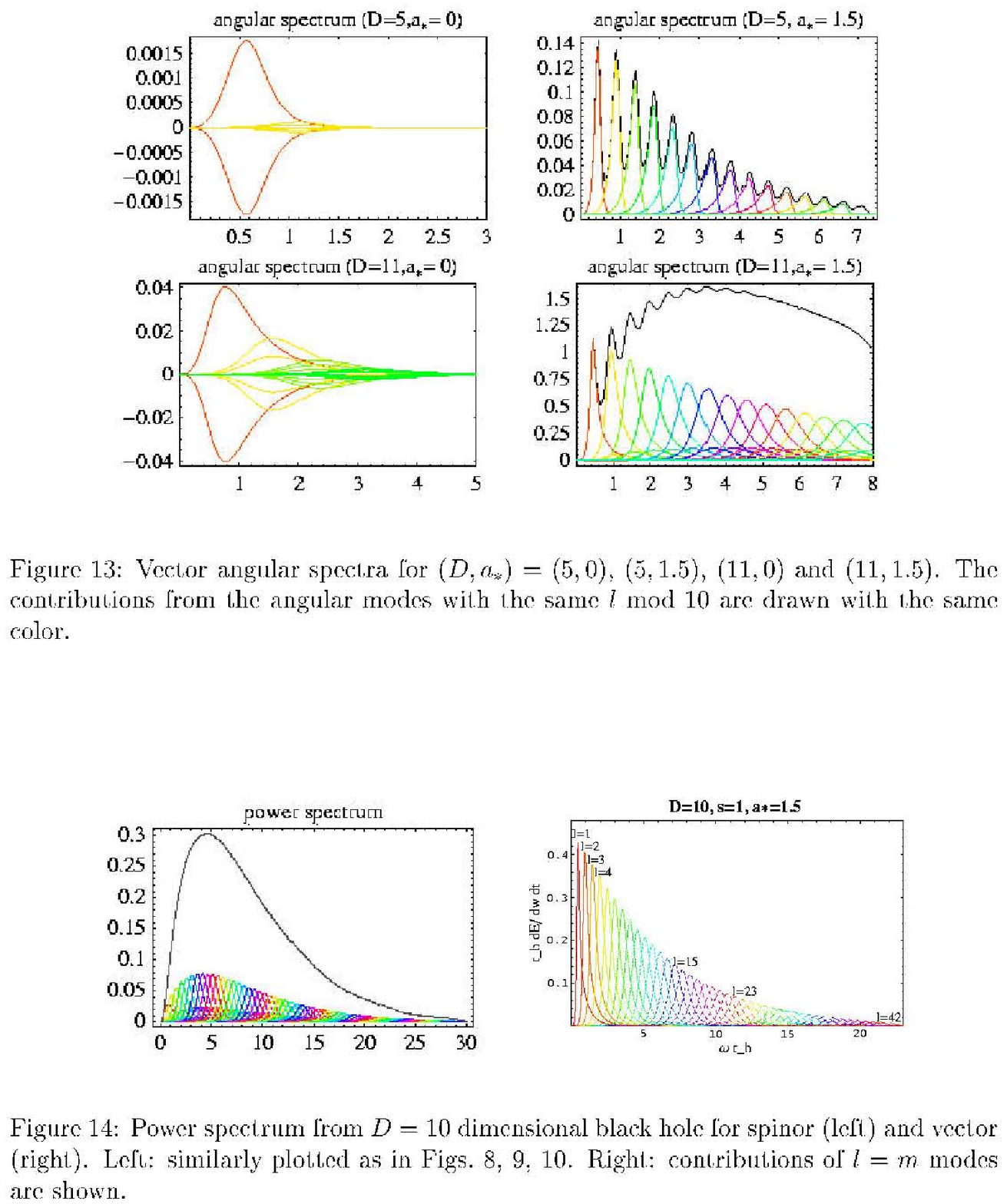}
\end{center}
\end{figure}
\begin{figure}[tbp]
\begin{center}
  \includegraphics{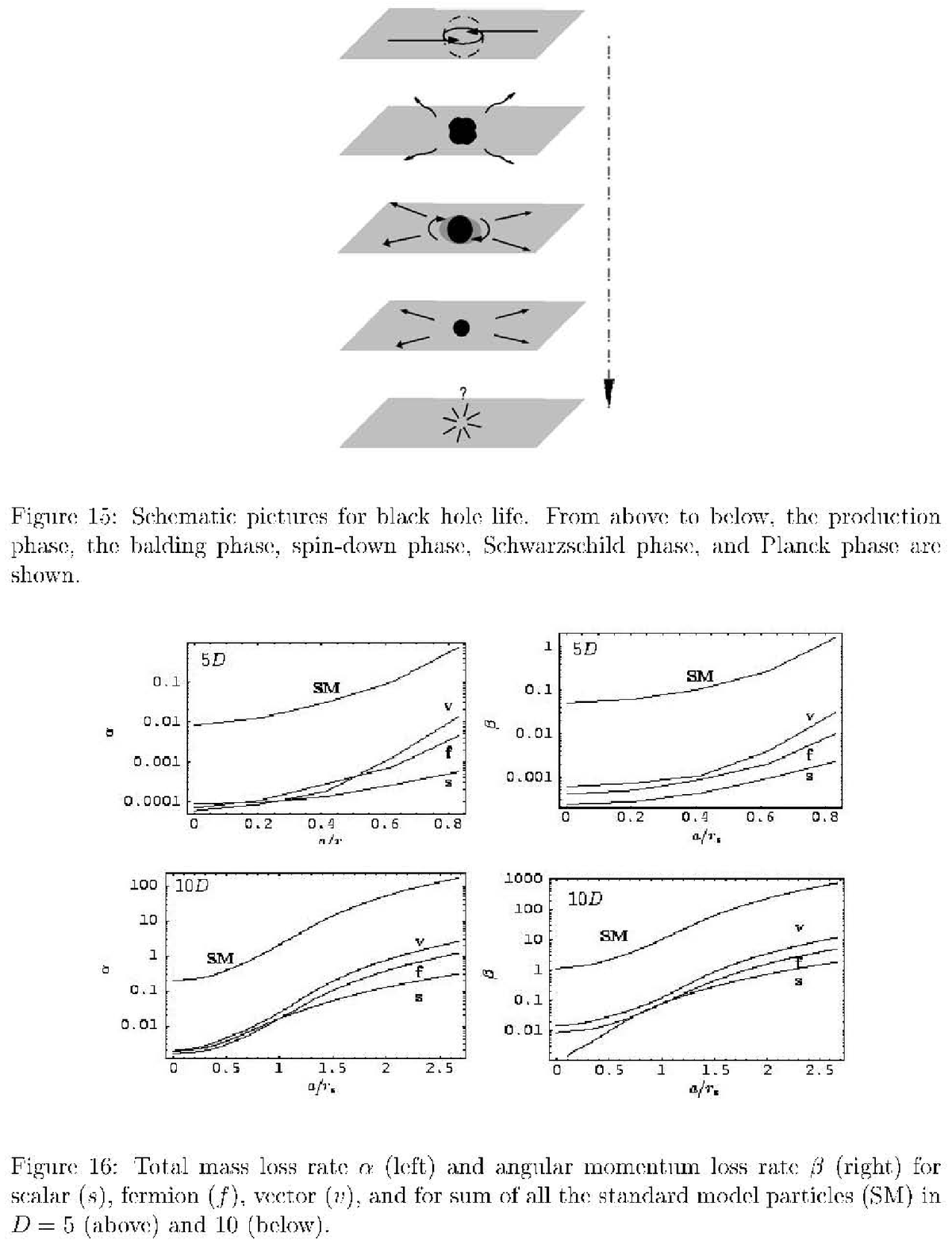}
\end{center}
\end{figure}

\end{document}